\documentclass[reprint,amsmath,amssymb,aps,prab,floatfix]{revtex4-2}
\usepackage[english] {babel}
\usepackage{amsfonts,graphicx,soul,natbib,mathrsfs}
\usepackage[euler]{textgreek}
\usepackage{epstopdf}
\usepackage{xcolor}
\usepackage{soul}
\usepackage{physics}
\usepackage{graphics}

\usepackage{dcolumn}
\usepackage{bm}

\usepackage[colorlinks=true]{hyperref}

\begin{document}

\title{On ``Efficiency versus Instability in plasma accelerators''}

\author{S. S. Baturin}%
\email{s.s.baturin@gmail.com}
\affiliation{Department of Electrical Engineering and Department of Physics, Northern Illinois University, DeKalb, IL 60115, USA}%

\date{\today}

\begin{abstract}
Recently in Ref.\cite{LBN} it was demonstrated that the efficiency of the energy transfer from the drive bunch to the witness bunch in the plasma wakefield accelerator has a limit due to the BBU instability of the witness bunch. It was stated that the efficiency-instability relation is universal and thus should be considered as a fundamental limit. In this note, we show that recent results on the short-range wakefields indicate that this relation should be modified and conclusions of Ref.\cite{LBN} should be reconsidered. In particular, we argue that the efficiency-instability relation produces only the lower bound for efficiency and thus does not produce a fundamental limit.      

\end{abstract}

\maketitle
\section{Brief statement}

It is suggested that the efficiency-instability relation derived in Ref.\cite{LBN} should be reconsidered as far as the derivation that was provided by the authors heavily relies on certain approximation to the short-range wakefields produced by the witness beam in a plasma cavity. As it is shown below, expressions for the wake potentials have to be modified to be consistent with the Ref.\cite{myPRL,mySTAB} that the authors relied on, but what is more important is that all equalities that involve wake potentials, namely Eq.(19), Eq.(20), Eq.(21), Eq.(22) and Eq.(23) of Ref.\cite{LBN} should be replaced with inequalities such that wake potential on the left-hand side is always less than the exact expression that the authors used. 

In Ref.\cite{mySTAB} it was shown that the expression for the wake potential used by the authors is the upper bound for the real wake potential in the case of accelerator structures and probably any net-neutral channel. Consequently, what the authors presented is not necessarily extendable to any wakefield accelerator, as the authors suggest in the text. For the case of the plasma bubble, recently in Ref.\cite{Stup_PB} G. Stupakov showed that expressions for the short-range wakefields in the plasma cavity used in the paper are the upper bounds of the real wake potentials.

Consequently, the final efficiency instability relation should be modified approximately as follows
\begin{align}
\label{eq:evi}
\eta_t \lesssim \frac{\eta_p^2}{4\left(1-\eta_p\right)},
\end{align}
where we use $\lesssim$ to indicate that the final form of this relation should be modified because the expression for the transverse wake potential should be updated for the case considered by the authors of Ref.\cite{LBN}. 

It turns out that Eq.(21) of Ref.\cite{LBN} for the transverse wake potential is inconsistent with Ref.\cite{mySTAB} and Panofsky-Wenzel theorem. The correct formula is presented below. Thus, the right-hand side of the Eq.\eqref{eq:evi} has to be modified to account for the new expression.

This comment is organized as follows: first, it is demonstrated that Eq.(17) of Ref.\cite{LBN} is not a known theorem, and in general case contradicts the results of Ref.\cite{myPRL} that the authors rely on. Next, it is demonstrated that the expression for the longitudinal and consequently transverse wake potential that is derived in Ref.\cite{myPRL} have to be considered as upper bounds, and consequently the relation between the efficiency and instability in the suggested model should be modified to resemble the inequality in Eq.\eqref{eq:evi}.

\section{\label{sec:intro}Short range wake field theorem}
One of the important steps in the authors' derivation is the application of the so-called short-range wake theorem. They reproduce it in the Eq.(17) of the Ref.\cite{LBN} 
\begin{align}
\label{eq:PWi}
W_{r}\approx \frac{2r_0}{r_b^2}\int\limits_0^{\xi}W_\parallel(s)ds.
\end{align} 
Here $r_0$ is the displacement of the beam from the center. In the notations of Ref.\cite{LBN} transverse wake potential is a scalar, so it is safe to assume that the authors refer to the radial part of the wake potential that kicks the beam towards the boundary thus we adopt the following notation $W_r/r_0\equiv W_\perp$. It is worth to mention that authors refer to $W_\perp$ and $W_\parallel$ as just wake potentials, so it is assumed that $W_\perp=W_\perp (r,r_0,\phi,\phi_0,s)$ as well as $W_\parallel=W_\parallel (r,r_0,\phi,\phi_0,s)$ 

To justify this equation, the authors refer to the Ref.\cite{myPRL} and references therein. It is worth noting that neither Ref.\cite{myPRL}, nor any of the reference cited in Ref.\cite{myPRL} contain this particular expression. In fact, Eq.\eqref{eq:PWi} is in contradiction with the result of Ref.\cite{myPRL} (see Eqs.(13), (14) and Table 1 of Ref.\cite{myPRL}) and overlaps with the result of Refs.\cite{myPRL,mySTAB} only in a very specific case.

Indeed, let's consider Panofsky-Wenzel theorem (see for example \cite{PW,Chao,Zotter})
\begin{align}
\label{eq:PW}
\nabla W_\parallel =\frac{\partial \mathbf{W}_\perp}{\partial \xi}.
\end{align} 
Here in contrast to what is introduced by the authors $\mathbf{W}_\perp$ is a vector and represents a full transverse wake potential as defined in \cite{Chao,Zotter}.  

In polar coordinates the nabla operator is given by
\begin{align}
\nabla=\left(\frac{\partial }{\partial r}, \frac{1}{r}\frac{\partial}{\partial \phi} \right),
\end{align}
with this Eq.\eqref{eq:PW} takes the form
\begin{align}
\label{eq:PWr}
&\frac{\partial W_\parallel}{\partial r}=\frac{\partial W_r}{\partial \xi}, \\
\label{eq:PWas}
&\frac{\partial W_\parallel}{r\partial \phi}=\frac{\partial W_\phi}{\partial \xi}.
\end{align}
Here $W_r$ and $W_\phi$ are the radial and azimuthal components of the transverse wake potential. If one considers a pencil beam, then the instability is caused by the radial component of the wake potential and Eq.\eqref{eq:PWas} could be omitted from consideration. 

We integrate Eq.\eqref{eq:PWr} and arrive at
\begin{align}
W_r=\int\limits_0^{\xi}\frac{\partial W_\parallel(s)}{\partial r}ds.
\end{align}  
By comparing Eq.\eqref{eq:PWi} with Eq.\eqref{eq:PWr} we conclude that for the Eq.\eqref{eq:PWi} to be consistent with the Panofsky-Wenzel theorem the following equality should hold
\begin{align}
\label{eq:ass}
 \frac{2r_0}{r_b^2}\int\limits_0^{\xi}W_\parallel(s)ds\approx\int\limits_0^{\xi}\frac{\partial W_\parallel(s)}{\partial r}ds.
\end{align} 
To proceed further we consider the case of a vacuum channel and retarding material that is the basis of the model from the Ref.\cite{myPRL}. First we set $\xi=0^{+}$ - a positive limit to zero, then Eq.\eqref{eq:ass} reduces to
\begin{align}
\label{eq:ass0}
 \frac{2r_0}{r_b^2}W_\parallel(0^{+},r)\approx\frac{\partial W_\parallel(0^+,r)}{\partial r}.
\end{align} 
The solution to the equation above reads
\begin{align}
\label{eq:cr}
W_\parallel(0^{+},r)\approx W_\parallel(0^{+},0)\exp\left[\frac{2rr_0}{r_b^2} \right].
\end{align}
We note that if Eq.\eqref{eq:PWi} is a general theorem, then the corollary Eq.\eqref{eq:cr} is in contradiction with the result of \cite{myPRL,mySTAB} where it was shown and illustrated by simulations that when the particle that generates the wake is at the point  $\omega_0=r_0\exp(i \phi_0)$, the expression for the longitudinal steady-state wake potential at the point $\omega=r\exp(i \phi)$ reads 
\begin{align}
\label{eq:cfm}
W_\parallel(0^{+},r)=\frac{4 L}{r_b^2}\mathrm{Re}\left[f^\prime(\omega,\omega_0)^* f^\prime(\omega_0,\omega_0)\right],    
\end{align}
where $L$ is the interaction length and $f$ is the conformal mapping function that maps the region of interest (vacuum channel cross-section) onto a unit disk such that the point $\omega_0$ corresponds to the center of that disk. It is worth noting that corollary Eq.\eqref{eq:cr} contradicts known results on a longitudinal wake potential in a resistive wall round pipe that could be found for example in Ref.\cite{Chao}.

Thus, equation Eq.\eqref{eq:cr} is not general and consequently the initial theorem Eq.\eqref{eq:PWi} is not general as well. The limited applicability of the theorem Eq.\eqref{eq:cr} calls into question the generality of the result that the authors claim.

It follows from Ref.\cite{mySTAB} that in the case when the channel is circular and the particle that generates the wakefield is point-like in transverse dimensions and displaces along radius by $r_0$ Eq.\eqref{eq:cfm2} at $\phi=\phi_0=0$ reduces to
\begin{align}
\label{eq:circc}
W_\parallel(0^+,r)=W_\parallel(0^{+},0)\frac{r_b^4}{(r_b^2-rr_0)^2}    
\end{align}
with $W_\parallel(0^{+},0)=\frac{4 L}{r_b^2}$.

It is apparent that even in the case of the cylindrical channel corollary of theorem Eq.\eqref{eq:PWi}, it is still inconsistent with the result of Ref.\cite{myPRL,mySTAB}. There is only one approximation $r=r_0$ and $r_0^2/r_b^2\sim0$ when both formulas match. We reiterate that this is true only at one specific point $r=r_0$ - the location of the bunch.

\section{Universality of the wake potential}
The second important assumption of the Ref.\cite{LBN} is that in the case when the bubble has a perfect, circular cross-section, the expression for $W_\parallel(\xi_1,0)$ for a point witness particle inside the bubble could still be calculated using the universal formula 
\begin{align}
\label{eq:loss}
W_\parallel(\xi_1)=\frac{4 L}{r_b(\xi_1)^2}.
\end{align}
Here $r_b$ is the bubble radius and $\xi_1$ is the longitudinal position of the witness particle inside the bubble.

It was mentioned in Ref.\cite{Bane} based on Ref.\cite{Stup_PB} that this expression might not be applicable in a plasma bubble. While in general, this expression is indeed not correct, as was shown Ref.\cite{Stup_PB}, under certain assumptions it still holds, but what is more important that it always holds if the $=$ sign is replaced with the $\leq$ sign. This fact for structures and hollow plasma channels explicitly follows from the Relativistic Gauss theorem \cite{myPRL} and for the case of the plasma bubble from the analysis of the Ref.\cite{Stup_PB}.
 
Indeed, we consider Eq.(30) of Ref.\cite{Stup_PB} (note that there is a misprint in the original paper \cite{Stup_pr})
\begin{align}
W_\parallel(\xi_1)=\frac{4 L}{\tilde{r}_b^2}\frac{\Delta \hat{E}_z(r_b)}{2/\tilde{r}_b+\Delta \hat{E}_z(r_b)}.
\end{align}
Now if the normalized bubble radius $\tilde{r}=k_p r_b\gg2$ that corresponds to the case when the real bubble radius is significantly larger then the plasma wavelength, the expression above could be decomposed into a series and one arrives at
\begin{align}
W_\parallel(\xi_1)\approx \frac{4L}{\tilde{r}_b^2}\left(1-\frac{2\Delta \hat{E}_z(r_b)}{\tilde{r}_b}\right).
\end{align}
As long as $\Delta \hat{E}_z(r_b)\geq0$, we conclude after converting everything back to CGS units that
\begin{align}
\label{eq:StupU}
W_\parallel(\xi_1)\leq \frac{4 L}{r_b(\xi_1)^2}.
\end{align}
From this simple consideration we see that Eq.\eqref{eq:loss} gives the upper bound for the longitudinal wake potential and converges to an exact solution in the limit of the bubble radius much large then the plasma wavelength. So we observe that Eq.\eqref{eq:loss} may overestimate the wake potential. This fact is explicitly illustrated in the end of the Ref.\cite{Stup_PB}, where a numerical comparison between Eq.\eqref{eq:loss} and the exact calculation is presented.

It is important to mention that the considerations of Ref.\cite{Stup_PB} have only one assumption, namely that the longitudinal beam size has to be significantly less than the plasma wavelength. The charge of the witness as well as the size of the bubble could be chosen arbitrarily.

Therefore, two main conclusions are to be made. Fist - the initial equation Eq.\eqref{eq:loss} in Ref.\cite{LBN} should be substituted with the inequality Eq.\eqref{eq:StupU}, and this inequality is universal. 

Second, we observe that we can slightly modify the formalism of Ref.\cite{mySTAB} to calculate the upper-bound for the longitudinal wake potential in the bubble of arbitrary shape, as now the bubble can be considered as a cavity with a curved boundary that is propagating with the beam. Consequently, 
\begin{align}
\label{eq:cfm2}
\frac{\partial W_\parallel(\xi,\omega)}{\partial \xi}\leq\frac{4 L \rho(\xi)}{r_b^2}\mathrm{Re}\left[f^\prime(\omega,\omega_0)^* f^\prime(\omega_0,\omega_0)\right].    
\end{align}
As it was shown in Ref.\cite{mySTAB} in the case when the channel is circular and the particle that generates the wakefield is point-like in transverse dimensions and displaced along radius by $r_0$ Eq.\eqref{eq:cfm2} reduces to
\begin{align}
\label{eq:circ}
\frac{\partial W_\parallel(\xi,r)}{\partial \xi}\leq\frac{4 L \rho(\xi)}{r_b^2}\frac{r_b^4}{(r_b^2-rr_0)^2}.    
\end{align}
Here we assumed $\phi=0$ and thus omitted dependence on the polar angle for  brevity. Differentiating Eq.\eqref{eq:circ} and combining the result with the Panofsky-Wenzel theorem \eqref{eq:PWr}, we arrive at
\begin{align}
\label{eq:trest}
W_r(\xi,r)\leq8L\int\limits_{\xi_1}^{\xi}ds\int\limits_{\xi_1}^{s}d\tilde{s}\frac{\rho(\tilde{s})r_b(\tilde{s})^2 r_0}{(r_b(\tilde{s})^2-rr_0)^3}.
\end{align}  
 If we assume $r=r_0$ and $r_0\ll r_b(\xi)$, then the equation above simplifies to
 \begin{align}
 \label{eq:trin}
W_r(\xi,r)\leq8L\int\limits_{\xi_1}^{\xi}ds\int\limits_{\xi_1}^{s}d\tilde{s}\frac{\rho(\tilde{s}) r_0}{r_b(\tilde{s})^4}.
 \end{align}   
We note that inequality above holds only for very small displacements. To get a proper estimate one should proceed with Eq.\eqref{eq:trest} because the condition $r_0\ll r_b(\xi)$ will not be satisfied near the end of the bubble.
 
\section{Discussion} 

The efficiency-instability relation is universal if the equality presented in Ref.\cite{LBN} is replaced with an inequality. This dramatically affects the conclusion of Ref.\cite{LBN} on fundamental limitations, as now the efficiency is not bounded from the top it is bounded from the bottom. There is no fundamental limitation.

Indeed if one consider definition of the instability parameter and a pencil beam with vanishing transverse size then
\begin{align}
    \eta_t=-\frac{F_t}{F_r}\leq \frac{Q}{2\pi n_0 |e| r_0}\frac{W_r(r_0,\zeta)}{L}
\end{align}
with $F_t$ - the deflecting Lorentz force of the wakefield and $F_r$ - the ion focusing force. One immediately observe that if $F_t$ is replaced by it's universal expression given by Eq.\eqref{eq:circ} then the right hand side will always be greater then $\eta_t$. As far as further derivation that authors preset is based on approximate evaluation of the right hand side the final connection between $\eta_t$ and $\eta_p$ is an inequality.   

The exact expression for this inequality relation should be rederived in order to be consistent with Refs.\cite{myPRL,mySTAB}. Namely, the expression for the transverse wake potential has to be modified according to Eq.\eqref{eq:trest}, or at least Eq.\eqref{eq:trin}. If this is accomplished, approximating the transverse wake potential by Eq.\eqref{eq:trest} or Eq.\eqref{eq:trin}, the resulting relation has to be an inequality of a type
\begin{align}
    \eta_t\leq f(\eta_p).
\end{align}
It produces a lower bound to the efficiency $\eta_p$ in the case when $f(x)$ is a polynomial (or when the polynomial is an upper estimate for the $f(x)$) of a degree $n\geq 1$. 
 
 It is important to note that modification of the bubble radius with an effective bubble radius $r_{\mathrm{eff}}=r_b+\alpha k_p$ as was suggested partially by the authors in Ref.\cite{LBN} and was considered in Ref.\cite{Stup_PB} is empirical as it contains a free constant $\alpha$ and thus could not be considered as a rigorously derived theoretical expression. In contrast, the use of the exact bubble radius is rigorous as follows from the considerations of Refs.\cite{Stup_PB,mySTAB} but this implies that universal expression for the wake potential is an inequality \eqref{eq:StupU}. 
 
 It is worth to mention as well that according to the Panofsky-Wenzel theorem and Eq.\eqref{eq:circ}, if $r_b$ is replaced with $r_{\mathrm {eff}}$ then in Eq.\eqref{eq:trest} and Eq.\eqref{eq:trin} exactly the same substitution have to take place. This prohibits the use of different $\alpha$ parameters for longitudinal and transverse wake potential.
 
The model and efficiency-instability relation as introduced by the authors could be used in certain cases as a good estimate, but it does not provide a solid and rigorous proof of a general fundamental limitation. While this limitation may indeed exist, it is still yet to be found and proven.

\newpage
  
\bibliographystyle{ieeetr}

\bibliography{eff_vs_inst}

\begin{thebibliography}{1}

\bibitem{LBN}
V.~Lebedev, A.~Burov, and S.~Nagaitsev, ``Efficiency versus instability in
  plasma accelerators,'' {\em Phys. Rev. Accel. Beams}, vol.~20, p.~121301, Dec
  2017.

\bibitem{myPRL}
S.~S. Baturin and A.~D. Kanareykin, ``Cherenkov radiation from short
  relativistic bunches: General approach,'' {\em Phys. Rev. Lett.}, vol.~113,
  p.~214801, Nov 2014.

\bibitem{mySTAB}
S.~S. Baturin and A.~D. Kanareykin, ``New method of calculating the wakefields
  of a point charge in a waveguide of arbitrary cross section,'' {\em Phys.
  Rev. Accel. Beams}, vol.~19, p.~051001, May 2016.

\bibitem{Stup_PB}
G.~Stupakov, ``Short-range wakefields generated in the blowout regime of
  plasma-wakefield acceleration,'' {\em Phys. Rev. Accel. Beams}, vol.~21,
  p.~041301, Apr 2018.

\bibitem{PW}
W.~K.~H. Panofsky and W.~A. Wenzel, ``Some considerations concerning the
  transverse deflection of charged particles in radio‐frequency fields,''
  {\em Review of Scientific Instruments}, vol.~27, no.~11, pp.~967--967, 1956.

\bibitem{Chao}
A.~Chao, {\em Physics of Collective Beam Instabilities in High Energy
  Accelerators}.
\newblock Wiley and Sons, New York, 1993.

\bibitem{Zotter}
B.~W. Zotter and S.~Kheifets, {\em Impedances and Wakes in High Energy Particle
  Accelerators}.
\newblock WORLD SCIENTIFIC, 1998.

\bibitem{Bane}
K.~Bane, ``The impedance of flat metallic plates with small corrugations,''
  {\em arXiv}, p.~1711.03201, 2017.

\bibitem{Stup_pr}
G.~Stupakov Private communications.

\end{thebibliography}

\end{document}